# Thermal transport evolution due to nanostructural transformations in Ga-doped indium-tin-oxide thin films

Alexandr Cocemasov[1], Vladimir Brinzari[1], Do-Gyeom Jeong[2], Ghenadii Korotcenkov[1], Sergiu Vatavu[3], Jong S. Lee[2] and Denis L. Nika[1,*]

[1]E. Pokatilov Laboratory of Physics and Engineering of Nanomaterials, Department of Physics and Engineering, Moldova State University, Chisinau, MD-2009, Republic of Moldova

[2]Laboratory for Spectroscopy of Condensed Matter Physics, Department of Physics and Photon Science, Gwangju Institute of Science and Technology, Gwangju 61005, Republic of Korea

[3]Physics of Semiconductors and Devices Laboratory, Department of Physics and Engineering, Moldova State University, Chisinau, MD-2009, Republic of Moldova

*Correspondence: dlnika@yahoo.com

**Abstract:** We report on a comprehensive theoretical and experimental investigation of thermal conductivity in indium-tin-oxide (ITO) thin films with various Ga concentrations (0-30 at. %) deposited by spray pyrolysis technique. X-Ray diffraction (XRD) and scanning electron microscopy have shown a structural transformation in the range 15–20 at. % Ga from the nanocrystalline to the amorphous phase. Room temperature femtosecond time domain thermoreflectance measurements showed nonlinear decrease of thermal conductivity in the range 2.0–0.5 Wm$^{-1}$K$^{-1}$ depending on Ga doping level. Comparing density functional theory calculations with XRD data it was found that Ga atoms substitute In atoms in the ITO nanocrystals retaining Ia-3 space group symmetry. The calculated phonon dispersion relations revealed that Ga doping leads to the appearance of hybridized metal atom vibrations with avoided-crossing behavior. These hybridized vibrations possess shortened mean free paths and are the main reason behind the thermal conductivity drop in nanocrystalline phase. An evolution from propagative to diffusive phonon thermal transport in ITO:Ga with 15-20 at. % of Ga was established. The suppressed thermal conductivity of ITO:Ga thin films deposited by spray pyrolysis may be crucial for their thermoelectric applications.

# 1. Introduction

Indium oxide based compounds and composites with high thermal stability have been intensively investigated as perspective semiconducting materials for high-temperature thermoelectric applications. Additives or supplement oxides of different metallic nature like Zn, Sn, Ge, Ga, Y, Nb, Ni etc. were used [1–7] in such compounds and composites to improve their thermoelectric and optical properties. Some of these materials retained the lattice structure of the host oxide at large amounts of additives. The specifics of $In_2O_3$ lattice is its complexity and large number of atoms in the primitive cell (40). As a consequence one could expect the reduction of phonon thermal conductivity $\kappa_{ph}$ due to the trapping of heat energy by numerous optical modes and flattening of their energy dispersions [8], leading to lower phonon group velocities. The structural complexity of Ia-3 space group impedes investigations of various defect complexes in $In_2O_3$ and their impact on electronic, phononic and thermoelectric properties. In Ref. [9] some of us has reported on theoretical investigations of electronic properties of $In_2O_3$ with Sn-, Ga- and O-based point defect complexes, employing density functional theory. It has been shown that defect complexes strongly influence the electronic band structure and position of indium and oxygen atoms [9]. Defect complex driven deviation in both atomic masses and crystal lattice strain results in enhancement of phonon scattering by point defects and may affect the lattice thermal conductivity. Impact of different defect complexes and structural vacancies on thermal properties of $In_2O_3$ requires additional investigations.

The strong improvement of specific electronic, thermal and thermoelectric properties could be reached when the material is in nanocrystalline form. Recent theoretical results demonstrated that enhanced electrical conductivity $\sigma$ and thermopower $S$ providing an elevated power factor $PF = \sigma S^2$ can be achieved in ITO nanofilms [10] due to the filtering effect [10,11] of low energy conduction electrons at grain boundaries. The $PF$ values of around 3 mW/m·K$^2$ were reported for thermally aged ITO films with an optimal Sn content [12,13]. Thorough independent experimental study [14] confirmed the presence of potential barrier in the vicinity of grain boundaries which are responsible for the filtering of electrons despite the degeneracy of ITO conduction band. ITO nanofilms have also demonstrated reduced values of thermal conductivity as compared with bulk ITO [15]. The drop of thermal conductivity was explained by strong phonon scattering on grains, reduction of electronic part $\kappa_{el}$ of thermal conductivity due to the filtering of low energy conduction electrons and porosity of ITO films [15]. Due to the geometric complexity of the nanograin network the trapping of heat carrying vibrational modes, leading to the reduction of phonon thermal conductivity $\kappa_{ph}$, is also possible. Reduction of thermal conductivity without degradation of electronic parameters ($\sigma$, $S$) radically improves the thermoelectric efficiency, determined by figure of merit $ZT = \sigma S^2 T/\kappa_{tot}$. Looking for compounds with low $\kappa_{tot}$ and high enough $\sigma$ and $S$ is a mainstream in thermoelectric material science.

Our experiments with ITO:Ga thin films deposited by spray pyrolysis have shown quite large values of *PF* [13]. However, an influence of deposition parameters (like as pyrolysis temperature) on electric and thermal properties of such films require further investigations and optimization. ITO films remain totally nanocrystalline up to 50 at. % Sn [12], while nanocrystalline to amorphous phase transition in ITO(Sn≤10 at. %):Ga-based films occurs at Ga content ≥ 20-30 at. % and may lead to a sharp drop in thermal conductivity.

Thereby, in this work we focus on experimental and theoretical investigations of thermal conductivity in ITO:Ga thin films with widely varying Ga content and almost constant amount of Sn ~6 at. %. Such concentration of Sn [12,13] is optimal in terms of *PF* and *ZT*. The experiments were performed by an ultrafast (femtosecond) laser-induced time domain thermoreflectance method (TDTR) [16,17]. It is a powerful and versatile technique for thermal properties investigations of a large variety of bulk and nanoscale systems. Within the many advantages of TDTR in comparison with conventional thermal conductivity measurements is an excellent spatial resolution at a length scale below tens of micrometers. TDTR also requires minimal sample preparation for the measurements. Using density functional theory (DFT) and linearized Boltzmann transport equation (BTE) for phonons we investigate in detail how changes in phonon energy spectra affects both phonon scattering mechanisms and lattice thermal conductivity in ITO:Ga films at different Ga concentrations.

The rest of the paper is organized as follows. In Section 2 we discuss sample preparation and characterization. Section 3 describes TDTR technique of thermal conductivity measurements. Our theoretical model of phonon modes and lattice thermal conductivity in ITO:Ga films is presented in Section 4. Discussions of results are provided in Section 5. Conclusions are given in Section 6.

## 2. Sample preparation and characterization

ITO films (6 at. % of Sn content in sprayed solution) were deposited at T=350-360 °C by spray pyrolysis method on polished silicon substrates (1x1 cm$^2$). Additionally, within the same deposition procedure several films were doped by Ga at various concentrations in the range up to 30 at. %. Hereinafter, under Ga concentration we refer to the following ratio: [Ga]/([In]+[Sn]+[Ga]) where corresponding partial concentrations of compound are indicated in parentheses. The sprayed solution was prepared as a mixture of 0.2M precursors of InCl$_3$, SnCl$_4$·5H$_2$O and GaCl$_3$ by dissolving in the dimethylformamide (DMF). The required film thickness around 200 nm was roughly monitored during the deposition of the film by its color and by sprayed solution volume and then was measured by F20 Filmetrics instrument. All experimental samples were annealed for a stabilization of film structural parameters at $T_{an}$ =500 ºC during half an hour. Other details of our preparation method are provided in Ref. [18].

The deposited films were characterized for their structural properties by an EMPYREAN XRD system from Malvern Panalytical using Cu Kα ($\lambda$=1.5405 Å) radiation within a diffraction angle range of 15–70° in θ/θ mode. The surface roughness was measured by atomic force microscopy (AFM) using a Park XE7 instrument. Average surface roughness of deposited films did not exceed 10 nm. Surface images of the films were obtained by scanning electron microscopy (SEM) using the Hitachi S-4700 instrument. The films were monitored by using conventional secondary electrons mode at an accelerating voltage of 10 kV. Figure 1(a-d) presents SEM images for ITO:Ga films with Ga content of 0, 5, 10, and 15 at. %. At Ga concentrations more than 15 at. % the films demonstrated totally amorphous structure. The XRD summary pattern for all samples with tested Ga contents shown in Figure 2 confirms this conclusion. The $In_2O_3$ characteristic peaks practically disappear at 20 at. % Ga. At the range 0-15 at. % of Ga XRD 2θ diffractograms show the existence of $In_2O_3$ nanocrystalline phase, although it does not imply the absence of finely dispersed $Ga_2O_3$ phase.

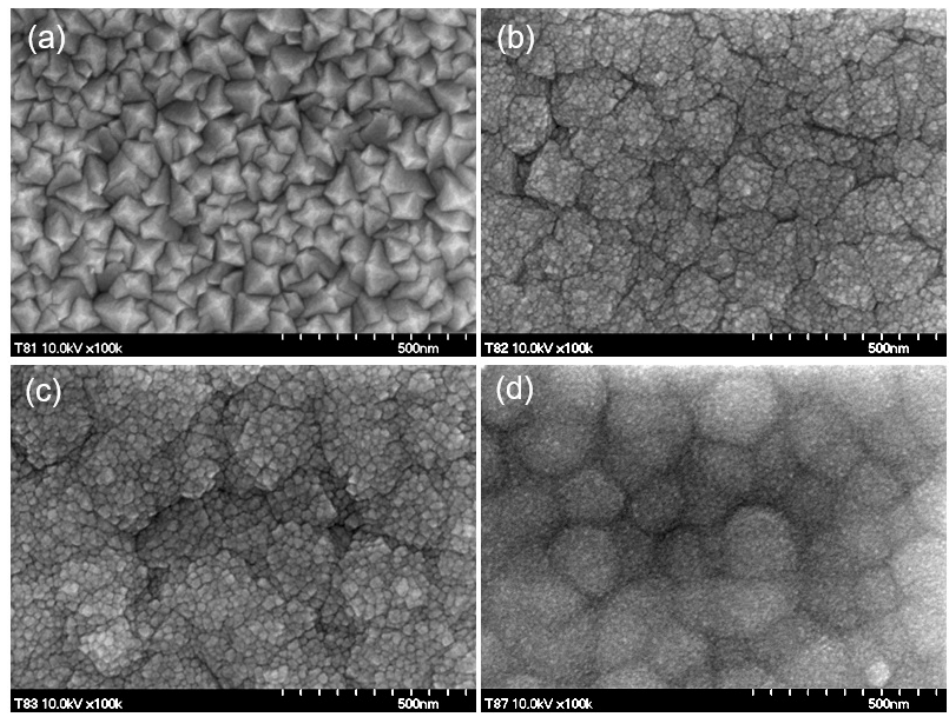

**Figure 1**(**a-d**). SEM images of ITO:Ga films with 0, 5, 10, and 15 at.% Ga.

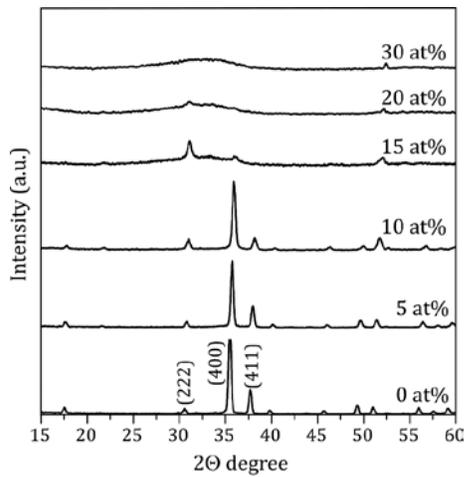

**Figure 2**. XRD intensity pattern for ITO:Ga films.

These two methods allow to compare polycrystalline effective grain sizes in cross-plane (extracted from XRD pattern by Scherrer formula) and in-plane (extracted from SEM images) directions, respectively. In Figure 3 we show corresponding values of average grain sizes versus Ga content in the measured films. The difference in grain sizes extracted from XRD and SEM patterns for pure ITO film reflects the anisotropy growth during deposition that is peculiar for large enough grains with surface faceting. The Ga additive manifests in smaller round-shaped nanocrystallites and in a tendency to their agglomeration. The position of dominated XRD $In_2O_3$ peaks and their shift with Ga content allowed us to calculate the changes in the lattice constant.

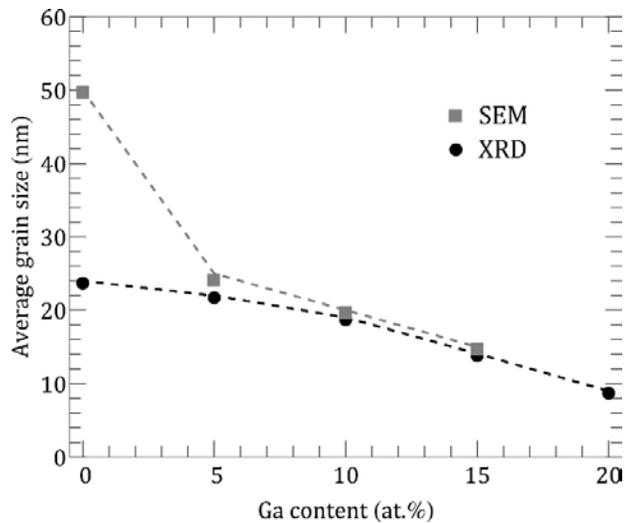

**Figure 3**. Average grain size in ITO:Ga thin films. Squares and circles denote SEM and XRD data, respectively.

## 3. TDTR measurements

The thermal transport properties were examined by femtosecond laser-induced time domain thermoreflectance method. An Al film was pre-deposited on the ITO:Ga films, and it was used as the thermal transducer of femtosecond laser pulses. The pump beam was modulated by an electro-optic modulator at 10 MHz and illuminated the transducer with a Gaussian radius of 5.9 μm. The pump-induced reflectivity change was monitored by the probe beam of which amplitude is collected with a fast amplified silicon detector connected to a lock-in amplifier. We obtained in-phase ($V_{in}$) and out-phase ($V_{out}$) profiles of the pump-beam-induced reflectivity variation of which phase is compared to the internal sinusoidal reference signal of the lock-in amplifier. We fitted the TDTR signal of $-V_{in}/V_{out}$ in a temporal range between 100 ps and 3.75 ns by considering the thermal transport model in multi-layered geometry based on the Fourier's law [16]. We set the boundary conductance across the Al-ITO interface and thermal conductivity of ITO films as two adjustable parameters. Heat capacities of Al ($C_{Al}$=2.42 Jcm$^{-3}$K$^{-1}$), Si (1.64 Jcm$^{-3}$K$^{-1}$) and ITO ($C_{ITO}$=2.78 Jcm$^{-3}$K$^{-1}$) were taken from the literature [19,20]. The thermal conductivity of Al transducer was obtained from the direct current electric conductivity by using a Wiedemann-Franz law. The Al film thickness $t_{Al}$ was estimated to have an average thickness of 100±3.6 nm by taking the position of an acoustic echo peak in the TDTR signal. The oscillating thermal wave cannot reach the bottom silicon substrate because of the very low thermal diffusivity of ITO:Ga thin films. The average oscillating heat penetration depth is estimated to be within 34-65% of ITO:Ga thickness. Because the lock-in amplifier independently captures and separates the oscillating thermal responses from steady-state heating, the thin films investigated here can be considered to be semi-infinite.

To check the validity and credibility of obtained thermal boundary conductance and thermal conductivity, we examined how the final modeling results are sensitive to those parameters in the given measurement configuration. The sensitivity is determined as the relative change of TDTR signal with respect to the relative change of the parameter of interest. We found that our TDTR results are highly sensitive to $\kappa_{ITO}$, but not to the other parameter $G_{Al-ITO}$. The latter is attributed to the very low thermal diffusivity of ITO samples so that the thermal gradient of oscillating heat is mostly given in ITO. Despite the low sensitivity to $G_{Al-ITO}$, we could successfully determine the quantity since the covariance between the two fitting parameters is small as about 0.2-0.5, except for the Ga content of 20 at. % having the value close to 0.6. In the error estimation, we considered the sensitivity results, uncertainties in the pre-parameters used in the thermal transport model, and the position-dependent inhomogeneity. After merging all these factors, we estimated the final error by minimizing Kullback-Leibler divergence [21]. The obtained parameters of $G_{Al-ITO}$ and $\kappa_{ITO}$ are presented in Figures. 4(a) and (b), respectively. At 20 at. % of Ga the error bar of $\kappa_{ITO}$ is relatively small, while that of $G_{Al-ITO}$ is much larger than error bars for other Ga content. The latter is attributed to the large covariance between $\kappa_{ITO}$ and $G_{Al-ITO}$ and to the high thermal resistivity of the amorphous phase, which allow us to determine $\kappa_{ITO}$ more reliably than $G_{Al-ITO}$.

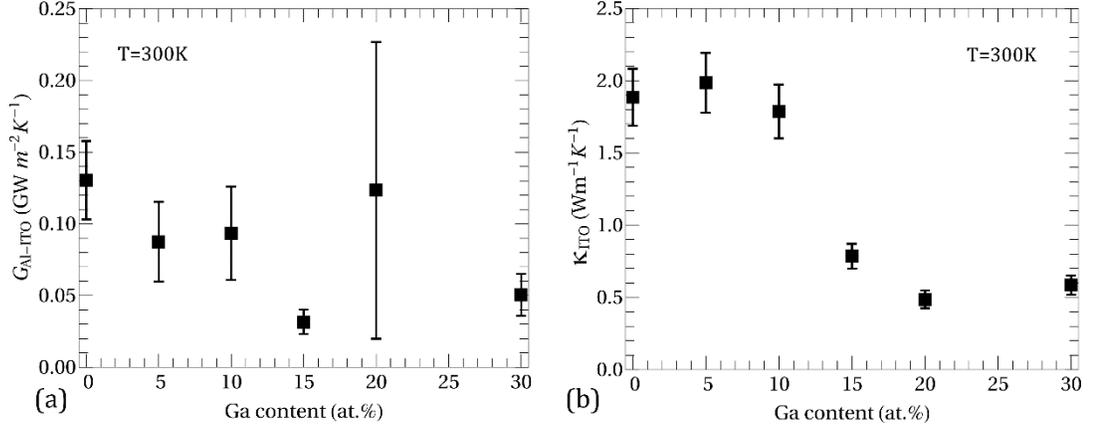

**Figure 4**. (**a**) Thermal boundary conductance across the Al-ITO interface GAl-ITO. (**b**) Thermal conductivity of ITO thin films $\kappa_{ITO}$.

## 4. Theoretical methods

All electronic calculations were performed within density functional theory formalism as implemented in the SIESTA code [22]. A generalized gradient approximation (GGA) for exchange-correlation functional of Perdew, Burke and Ernzerhof (PBEsol) [23] was used. Scalar-relativistic norm-conserving pseudopotentials for core states [24] generated with PseudoDojo [25] and double-zeta basis set for localized atomic orbitals were employed, with the following valence configurations for atoms: In $4s^2 4p^6 4d^{10} 5s^2 5p^1$, Sn $4s^2 4p^6 4d^{10} 5s^2 5p^2$, Ga $3s^2 3p^6 3d^{10} 4s^2 4p^1$ and O $2s^2 2p^4$. A real-space mesh with energy cutoff of 300 Ry and 3x3x3 reciprocal-space mesh of the Monkhorst-Pack type [26] were found sufficient to perform structure optimizations and calculate the second-order interatomic force constants. The force and stress convergence criteria were set to 0.001 eV/Å and 0.005 GPa, respectively. For relaxed $In_2O_3$ with 40-atom primitive cell ($In_{16}O_{24}$) we have obtained a lattice constant of 10.048 Å and volumetric density of 7.27 g/cm$^3$, which are close to experimental values of 10.117 Å [27] and 7.18 g/cm$^3$ [28].

The ITO with various Ga doping concentrations was simulated by adding (removing) a certain number of Ga (In) atoms in a 40-atom primitive cell of bixbyite crystal structure, an example of which is shown in Figure 5. In all cells one indium atom at *b*-site was replaced by a tin atom, which corresponds to ITO with 6.25 at. % Sn-doping. The harmonic force constants were calculated using the finite difference method as implemented in the PHONOPY package [29]. The displacement length of each atom from its equilibrium position was set to 0.015 Å.

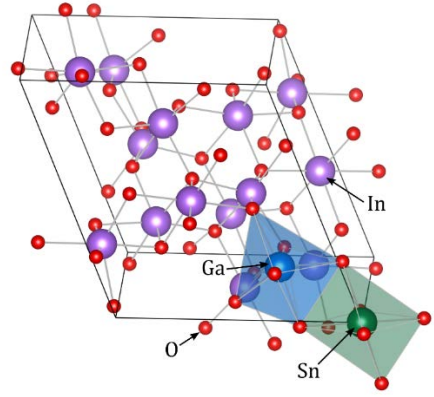

**Figure 5**. Example of ITO:Ga crystal structure. Two octahedra denote Sn (green) and Ga (blue) substitutional defects surrounded by six nearest oxygen atoms.

Once the interatomic force constants were obtained, the dynamical matrix was constructed and diagonalized, thus obtaining phonon dispersions $\omega_s(\boldsymbol{q})$, where *s* is phonon branch number and $\boldsymbol{q}$ is phonon wavevector. Based on the phonon dispersion data the phonon thermal conductivity was calculated using the linearized Boltzmann transport equation:

$$\kappa_{ph} = \frac{1}{NV}\sum_{\boldsymbol{q},s} \hbar\omega_s(\boldsymbol{q})v_{s,x}^2(\boldsymbol{q})\tau_s(\boldsymbol{q})\frac{\partial f}{\partial T}. \tag{1}$$

In Equation 1 summation is performed over entire Brillouin zone mesh and all phonon branches, *T* is the temperature, *N* is the number of *q*-mesh points, *V* is the unit cell volume, $\omega$ and $v_x = d\omega/dq_x$ are the phonon frequency and group velocity along the thermal gradient, $\tau$ is the phonon lifetime and *f* is the Bose-Einstein distribution function. Converged results for thermal conductivity were achieved for a dense 20x20x20 *q*-point mesh (*N*=8000 points). It is implied in Equation (1) that thermal flux is directed along the temperature gradient, that is along *X* Cartesian axis. The choice of the latter is formal, since thermal conductivity tensor in bixbyite crystals is isotropic.

In case of crystalline (nanocrystalline) structures the phonon scattering rate was calculated according to the Matthiessen's rule:

$$1/\tau = 1/\tau_U + 1/\tau_{GB}, \tag{2}$$

where $1/\tau_U = BT\omega_s^2(\boldsymbol{q})e^{-1/T}$ is the three-phonon Umklapp scattering rate [30] and $1/\tau_{GB} = v_s(\boldsymbol{q})/l_G$ is the phonon scattering rate on grain (i.e. nanocrystal) boundaries with the average grain size $l_G$. The scattering on grains is assumed to be diffusive, so all incident phonons completely lose their momenta after they reach the grain boundary. Parameter $B=10^{-18}$ s/K of the Umklapp scattering was found from the comparison between calculated and experimental [7] thermal conductivities of pure $In_2O_3$, i.e. without phonon scattering on grains.

In case of amorphous structures, the dominant mechanism of heat transport is diffusion between localized atomic vibrations [31], in contrast to the wave-like phonon transport in crystals. In order to model the diffusive character of thermal transport in amorphous ITO:Ga we employ here an approach developed by some of us in Ref. [32]. Comparing equations for lattice thermal conductivity within linearized BTE (see Equation (1)) and within a random-walk diffusion theory of heat conduction [33], and can obtain the following expression for the rate of diffusion of lattice vibrations in an amorphous material:

$$1/\tau = \frac{3\pi v_s^2(\boldsymbol{q})\omega_s(\boldsymbol{q})}{V_0^{2/3}\langle\omega\rangle^2}, \quad (3)$$

where $V_0$ is the volume per atom and $\langle\omega\rangle$ is the mean vibrational frequency, which we have estimated from the condition that $\langle\omega\rangle$ is the frequency at which integrated phonon density of states (DOS) equals one-half: $\int_0^{\langle\omega\rangle} g(\omega)d\omega = 1/2$. In this way we avoided using any free parameters in our thermal conductivity model for ITO:Ga films.

The electronic part $\kappa_{el}$ was estimated according to the filtering model [10,11,34]. Necessary parameters of ITO used in the calculations were taken from [35]. Electron scattering on polar optical phonons and ionized impurities were taken into consideration. The total thermal conductivity was obtained as a sum of the phononic and electronic parts: $\kappa_{tot} = \kappa_{ph} + \kappa_{el}$. Also, the porosity is an inevitable structural factor for spray-pyrolyzed structures. The effect of porosity on thermal conductivity of ITO:Ga was accounted according to the effective medium theory [36,37]. The porosity of our films was estimated in the range 25-35% by comparing the real refractive index of the films measured by laser ellipsometry with the same one for the bulk material [38]. This porosity is within a reasonable range, since 44% porosity of $In_2O_3$ [39] was reported for samples synthesized at 100 °C. We note here, that we do not clarify an influence of pore ordering and their geometry on $\kappa_{tot}$, since currently we have no technological control over these factors. This question requires an additional investigations and remains to be addressed in the future. Thereby, for all calculated structures the porosity was constant and fixed to 30% to facilitate the theoretical analysis and rather to focus on understanding how changes in phonon dispersion and scattering mechanisms affect the thermal transport at various Ga concentrations. It is worth noting, that higher porosity means lower thermal conductivity, thus providing another way for phonon engineering [40–43] and tuning the thermoelectric performance of indium oxide based material.

## 5. Results and Discussions

For accurate modeling of the thermal conductivity in our structures we should determine which position of the ITO lattice tend to occupy Ga atoms. For this purpose, we have extracted the lattice constants at different Ga concentrations from our XRD measurements and compared them with theoretical values calculated from DFT relaxation procedure. Figure 6 shows the evolution of the ITO lattice constant as a function of Ga content.

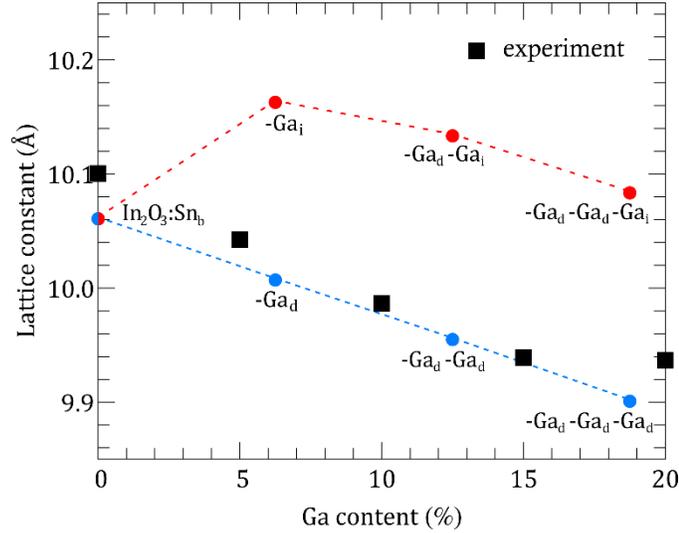

**Figure 6**. Lattice constant of ITO:Ga as a function of Ga content. Squares denote XRD data. Red and blue circles denote theoretical calculations with Ga atoms placed in different positions (interstitial c-site and substitutional d-site).

According to our previous work on $In_2O_3$ point defect formation energies [9], there are in general two possible scenarios for Ga atoms. In the first scenario all Ga atoms occupy substitutional positions (Figure 6, blue circles), which are the most energetically favorable. Here the theory predicts the compression of ITO crystal structure with a linear decrease of lattice constant given by the law: $a(x) = a(0) - 0.0085x$, where $a(0) = 10.06$ Å. In the second scenario, which is less likely from energetical point of view, Ga atom is placed at an intersite ($c$-site, i.e. structural vacancy, specific for Ia-3 space group). In this case calculations predict an expansion of the crystal structure with an increase of the lattice constant from 10.06 Å at 0 at. % Ga to 10.16 Å at 6.25 at. % Ga. However, introduction of additional Ga atoms in substitutional positions results again in the lattice compression (Figure 6, red circles). According to experimental data (Figure 6, squares) the lattice constant decreases when Ga content increases from 0 at. % to 15 at. %, thus supporting the first theoretical scenario with all Ga atoms forming substitutional point defects. Our study of ITO:Ga system demonstrates that within the given technological conditions the bixbyite lattice structure is maintained until a quarter of In atoms is substituted by guest atoms and no secondary oxide phases ($SnO_2$, $Ga_2O_3$) appear with non-bixbyite crystallographic symmetries, that confirms the high stability of cubic $In_2O_3$ structure. On the other hand, such substitution value (~0.25) also means a high solubility limit of Ga atoms in ITO system. Further increase of Ga doping dramatically changes the crystal structure: XRD and SEM measurements (see Figures 1-2) show that the structures become amorphous and lose the long-range atomic order. Nevertheless, it means that formation of In-Sn-Ga-O compound occurs even with increased Ga content in the deposited precursor. Otherwise, traces of the bixbyite structure should have been preserved on the XRD spectra.

Figure 7a shows the room temperature total thermal conductivity in ITO:Ga films as a function of Ga content. Squares denote experimental data, while circles represent theoretical calculations of $\kappa_{tot}$ with Ga atoms in substitutional positions. Calculation of $\kappa_{el}$ within filtering model [10,11,34] is also presented for comparison (triangles). Intending to elucidate how the changes in crystalline structure affect the thermal conductivity at different Ga concentrations, we have performed two calculations of $\kappa_{ph}$: (i) with phonon lifetime for crystalline films according to Equation (2) (blue circles in Figure 7a) and (ii) with phonon lifetime for amorphous films according to Equation (3) (green circles in Figure 7a). In general, the experimental results reveal a decreasing trend in $\kappa_{tot}$ for increasing Ga doping up to 15-20 at. %, which reasonably agrees with calculations employing phonon lifetimes for crystalline films. As the level of disorder increases further (20 at. % and 30 at. % Ga) the thermal conductivity saturates around value ~ 0.5-0.6 $Wm^{-1}K^{-1}$, which is close to the theoretical predictions with phonon lifetimes for the amorphous case. Thus, our theoretical results suggest that in the range 15-20 at. % Ga there is an evolution from propagative (i.e. crystalline) to diffusive (i.e. amorphous) phonon thermal transport in ITO:Ga films, which is in a good agreement with XRD and SEM measurements showing crystal-to-amorphous structural transition.

At the same time our estimation of the electronic thermal conductivity demonstrates reduction of $\kappa_{el}$ from 0.28 $Wm^{-1}K^{-1}$ at 0 at. % Ga down to 0.19 $Wm^{-1}K^{-1}$ at 18.75 at. % Ga. Amplification of electron scattering arises from reduction of the grain size with the rise of Ga content. The electron concentration in the conduction band remains unchanged because of isovalent nature of Ga impurity. The obtained values of $\kappa_{el}$ are in the same range as $\kappa_{ph}$ in amorphous calculation, while they are several times lower than $\kappa_{ph}$ in crystalline calculation. It is clear from Figure 7a that decreasing of total thermal conductivity of ITO:Ga films as Ga concentration increases up to 15-20 at. % is mainly due to suppression of the phononic contribution. Specifically, there are two responsible factors: (i) reduction of the average grain size $l_G$ (see Figure 3) with the corresponding increase of the phonon scattering rate and (ii) modification of the phonon energy spectra accompanied by diminution of the average group velocity of phonons (see Figure 8b below). These are the main factors behind the thermal conductivity drop in the nanocrystalline ITO:Ga films.

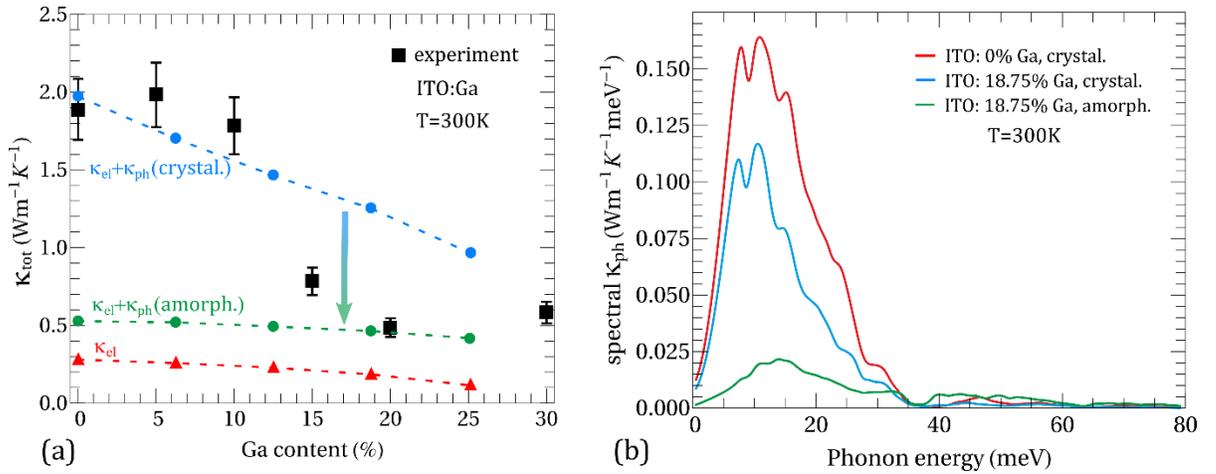

**Figure 7**. (**a**) Total thermal conductivity of ITO:Ga films at 300 K as a function of Ga content. Squares represent experimental data, while circles denote theoretical calculations with phonon lifetimes for crystalline (blue circles) and amorphous (green circles) materials. Electronic thermal conductivity is denoted with red triangles. (**b**) Spectral phonon thermal conductivity in ITO with 0 at. % Ga (red lines) and 18.75 at. % Ga (blue and green lines) as a function of phonon energy. Calculations with "crystalline" and "amorphous" phonon lifetimes are presented.

A useful quantity that reveals the contribution of phonons at various energies into heat conduction is the spectral phonon thermal conductivity $\kappa_{ph}(E)$, determined by: $\kappa_{ph} = \int \kappa_{ph}(E)dE$. Figure 7b shows $\kappa_{ph}(E)$ in crystalline and amorphous calculations for ITO with 0 at. % Ga (red line) and 18.75 at. % Ga (blue and green lines). The largest contribution to $\kappa_{ph}$ comes from phonons with energies in the range 0-35 meV. The contribution of higher energy modes >40 meV in the amorphous case (green line) is ~25% of $\kappa_{ph}$, while it is almost negligible in the crystalline case (red and blue lines) due to an enhanced phonon-phonon Umklapp scattering characterized by a quadratic dependence on phonon frequency: $1/\tau_U \sim \omega^2$. Moreover, it is obviously seen that suppression of thermal conductivity with increasing Ga content in the crystalline case is due to the weaker contribution from low-energy (0-35 meV) vibrations. In case of amorphous films (>15 at. % Ga) the $\kappa_{ph}(E)$ dependence is almost insensitive to the amount of Ga since according to Equations (1) and (3) $\kappa_{ph}(E) \sim \langle \omega \rangle$, while mean vibrational frequency $\langle \omega \rangle$ deviates very weakly for different Ga concentrations e.g. $\hbar \langle \omega \rangle = 46$ meV at 0 at. % Ga and $\hbar \langle \omega \rangle = 45$ meV at 18.75 at. % Ga.

We now turn to a more detailed analysis of the phonon properties and associated effects behind the thermal conductivity suppression in our ITO:Ga films. The phonon dispersion and projected density of states (PDOS) in ITO with 18.75 at. % Ga is shown in Figure 8a. The PDOS curves are presented per one atom of the respective specimen. The maximum energy of optic phonons is ~80 meV and is only weakly dependent on Ga doping. Three acoustic branches (light green curves) - one longitudinal acoustic (LA) and two transversal acoustic (TA) demonstrate a linear dispersion near the Brillouin zone center, with sound velocities: $v_{LA}$=4.7 km/s and $v_{TA}$=2.9 km/s. At 0 at. % Ga $v_{LA}$=4.8 km/s and $v_{TA}$=3.0 km/s, so one can conclude that rise of Ga content results in a weak softening of the acoustic branches.

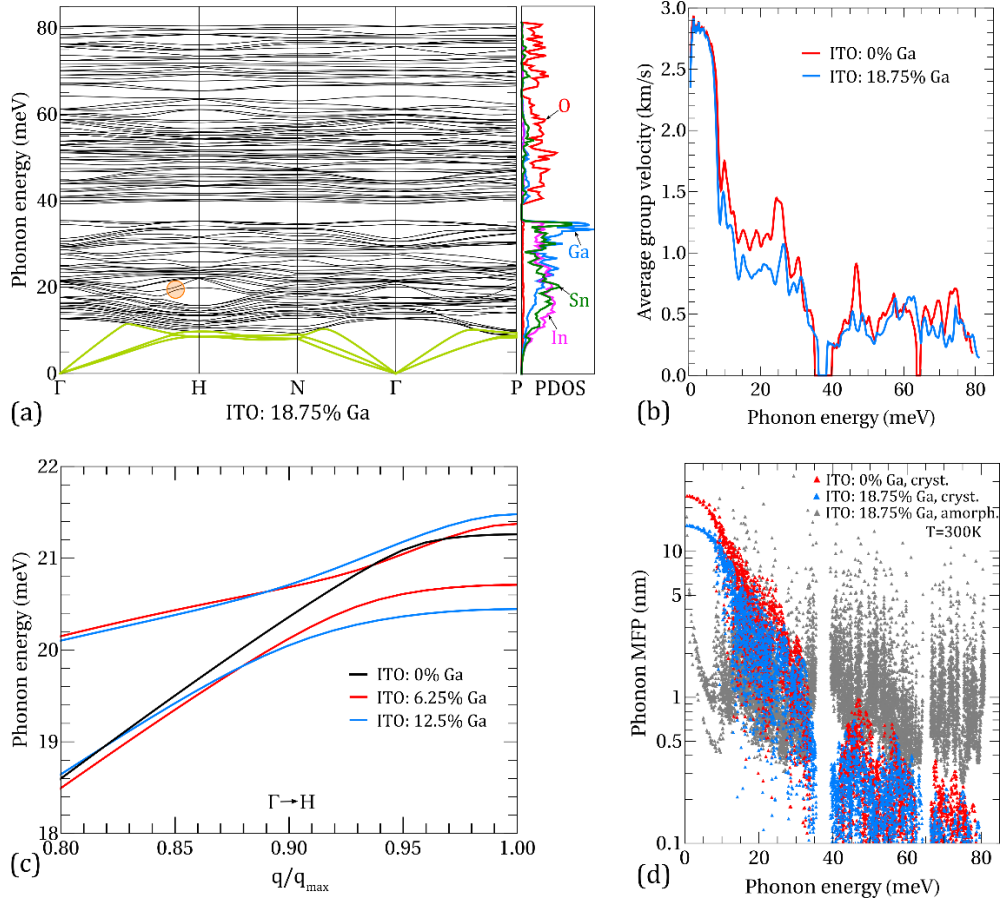

**Figure 8**. (**a**) Phonon dispersion and PDOS in ITO with 18.75 at. % Ga. The green curves in phonon spectra highlight acoustic branches. The orange circle designates the region with a clear avoided crossing. (**b**) Average phonon group velocity as a function of phonon energy in ITO with 0 at. % Ga (red line) and 18.75 at. % Ga (blue line). (**c**) Avoided-crossing behavior of phonon dispersion in ITO:Ga. (**d**) Phonon mean free paths at 300 K.

According to the dispersion and PDOS data it is natural to split the whole energy range into the lower part (0-35 meV), which is dominated by the vibrations of the metal atoms and the upper part (>40 meV) in which the vibrations of lighter oxygen atoms determine the vibrational spectrum. These two parts are separated by an energy gap in the range ~35-40 meV, which slightly narrows as Ga concentration increases. The low-energy part of the spectra is characterized by hybridization (mixing) of metal-type vibrational modes - mixed In/Sn, In/Ga and Sn/Ga modes. These hybridized modes are accompanied by the appearance of multiple avoided-crossing points throughout the Brillouin zone (see Figure 8c for an example), resulting in the flattening of the dispersion law and a corresponding decrease of the average group velocity at phonon energies between 10 and 30 meV (see Figure 8b). This effect is more pronounced for higher Ga concentrations and it plays an important role in the suppression of the phonon heat conduction in nanocrystalline ITO:Ga films. The similar avoided-crossing behavior was found in "host-guest" type materials such as $Ba_8Ga_{16}Ge_{30}$ clathrate [44] and $YbFe_4Sb_{12}$ skutterudite [45]. An analytical model describing the phonon dispersion relation of host-guest lattices with heavy guest atoms was recently proposed in Ref. [46].

It is insightful to separate between the intrinsic and extrinsic factors influencing the $\kappa_{ph}$ of ITO:Ga at different Ga-doping. The intrinsic element is the modification of phonon dispersion and group velocities, related to the changes in the atomic composition and interatomic forces. From the extrinsic factors, we can distinguish the variation in average grain size, which directly affects the phonon scattering rate on the grain boundaries. It should be noted, that extrinsic factors usually could be tailored within a certain limit by the technological conditions, thus providing additional way for managing the thermal flux in ITO-based structures. In order to quantify the influence of the above factors on $\kappa_{ph}$ we show in Figure 8d the mean free paths (MFPs) of phonons calculated taking into account Umklapp processes and phonon scattering on the grain boundaries (red and blue triangles). It can be seen that MFP of low-energy phonons (<10meV) in nanocrystalline ITO:Ga is limited merely by the scattering on grains with an average grain size $l_G$=24nm for 0 at. % Ga and $l_G$=15nm for 18.75 at. % Ga, as was determined from our XRD measurements presented in Figure 3. For higher energies, the phonon-phonon Umklapp scattering prevails and the phonon MFP quickly drops. Nevertheless, according to the Bose-Einstein distribution the population factors of the low-energy modes are smaller than of higher energy modes, as a result the overall contribution to the heat flux from the energy range 0-10 meV does not exceed 30% as determined from the spectral $\kappa_{ph}$ data (see Figure 7b). At the same time, phonons with energies 10-35 meV are the main heat carriers at room temperature transferring more than 2/3 of the total thermal flux. Therefore, the drop in thermal conductivity of nanocrystalline ITO:Ga films as the Ga content increases is primarily due to the redistribution of the vibrational spectrum in this energy range, leading to diminution of the average group velocity and MFPs of phonons. Finally, in the upper part of the spectra (>40 meV) all modes possess MFPs below 1 nm owing to the low group velocities and enhanced Umklapp scattering, resulting in an almost negligible contribution to the thermal conductivity as follows from $\kappa_{ph}(E)$ on Figure 7b. In contrast, the phonon MFPs in an amorphous calculation (gray triangles in Figure 8d) are almost uniformly distributed over entire energy range and this picture is independent of Ga content. It explains the relatively high contribution of high-energy phonon modes in the amorphous case as well as the much weaker dependence of phonon thermal conductivity on Ga amount in comparison with nanocrystalline ITO:Ga films.

## 6. Conclusions

Thermal properties of ITO thin films with Ga concentrations up to 30 at. % deposited by spray pyrolysis were investigated both experimentally and theoretically. XRD and SEM spectra showed the existence of $In_2O_3$ nanocrystalline phase in the range 0-15 at. % Ga, while for higher Ga concentrations the films demonstrated an amorphous structure. The incorporation of Ga atoms into the host lattice during the growth in given technological conditions is an important finding regarding the high Ga solubility limit in ITO. TDTR measurements revealed a decreasing of room temperature thermal conductivity with increase of Ga-content from 2.0 $Wm^{-1}K^{-1}$ at 5 at. % Ga down to 0.8 $Wm^{-1}K^{-1}$ at 15 at. % Ga, while for higher Ga concentrations thermal conductivity saturates ~0.5-0.6 $Wm^{-1}K^{-1}$. Theoretical modelling within DFT and BTE approaches demonstrated that this drop of thermal conductivity is primarily due to the redistribution of the metal atoms vibrations in the energy range 10-35 meV, accompanied by the appearance of multiple avoided-crossing modes throughout the Brillouin zone with decreased group velocity and shortened phonon MFPs. In the range 15-20 at. % Ga both experimental and theoretical results confirm the evolution from propagative to diffusive phonon thermal transport in ITO:Ga nanofilms. The obtained results demonstrate the practical possibility of engineering the thermal conductivity in ITO films by Ga doping and may lead to their thermoelectric applications.

**Author Contributions:** Conceptualization, V.B., D.G.J. and D.L.N.; software, A.C. and D.G.J.; validation, V.B., G.K., S.V. J.S.L. and D.L.N.; visualization, A.C. and D.G.J.; writing—original draft preparation, A.C., V.B. and D.G.J.; writing—review and editing, V.B., G.K., S.V., J.S.L. and D.L.N.; funding acquisition and project administration, J.S.L. and D.L.N. All authors have read and agreed to the published version of the manuscript.

**Funding:** A.C., V.B., G.K. and D.N. acknowledge the financial support from Moldova State Project no. 20.80009.5007.02. S.V. acknowledges the financial support from Moldova State Project no. 20.80009.5007.12. This work was supported by the Basic Science Research Program through the National Research Foundation of Korea (NRF) funded by the Ministry of Science, ICT & Future Planning (No. 2018R1A2B2005331).

**Conflicts of Interest:** The authors declare no conflict of interest. The funders had no role in the design of the study; in the collection, analyses, or interpretation of data; in the writing of the manuscript, or in the decision to publish the results.